\documentstyle[eqsecnum,pra,aps]{revtex}
\begin{document}
\title{A generalized Pancharatnam geometric phase formula for three
level quantum systems}
\author{Arvind\cite{email}} 
\address{Department of Physics\\
Indian Institute of Science,  Bangalore - 560 012, India}
\author{ K.~S.~Mallesh}
\address{Department of Studies in Physics\\ University of Mysore,
Mysore 570-006, India}
\author{N. Mukunda\cite{jncasr}} 
\address{Center for Theoretical Studies and Department of Physics\\
Indian Institute of Science,  Bangalore - 560 012, India}
\date{\today}
\maketitle
\begin{abstract}
We describe a recently developed generalisation of the 
Poincar$\stackrel{\prime}{\rm e}$ sphere 
method, to represent pure states of a three-level quantum system in a
convenient geometrical manner. The construction depends on the properties
of the group $SU(3)\/$ and its generators in the defining representation,
and uses geometrical objects and operations in an eight dimensional real
Euclidean space. This construction is then used to develop a generalisation
of the well known Pancharatnam geometric phase formula, for evolution of
a three-level system along a geodesic triangle in state space.
\end{abstract}
\section{Introduction}
\setcounter{equation}{0}
Sometime after the discovery by Berry in 1984 of the quantum
mechanical geometric phase in the framework of cyclic adiabatic 
evolution~\cite{bary-1}, Ramaseshan and Nityananda pointed out in an important
paper~\cite{rajaram} that as early as 1956 Pancharatnam  had put forward closely 
related ideas in the context of polarisation optics~\cite{pancharatnam}. Subsequently
Berry himself analysed the significance of Pancharatnam's work in the
light of later developments~\cite{bary-2}. The more recent quantum kinematic
approach to the geometric phase brings out in fully transparent
fashion the way in which a phase essentially identified by
Pancharatnam is one of the two basic ingredients involved in the
very definition of the quantum geometric phase, the other being the
so called dynamical phase~\cite{simon-mukunda-1}.  

Pancharatnam's work made essential use of the 
Poincar$\stackrel{\prime}{\rm e}$ sphere
representation for the manifold of pure polarisation states of a
plane electromagnetic wave~\cite{born-wolf}. As is well known, diametrically opposite
points on the 
Poincar$\stackrel{\prime}{\rm e}$ sphere correspond to mutually orthogonal
polarisation states incapable of interfering with one another.
For two states of polarisation not mutually orthogonal in this sense 
Pancharatnam introduced a physically motivated convention or rule
which would tell us when these two states are `` in phase '' , i.e.,
capable of interfering constructively to the maximum possible extent.
More precisely, this relationship is defined at the level of field
amplitudes mapping onto given points on the 
Poincar$\stackrel{\prime}{\rm e}$ sphere.
He then went on to show that this relation of being `` in phase ''
{\em is not transitive}. That is, if we take three polarisation
states $A, B, C\/$ on the 
Poincar$\stackrel{\prime}{\rm e}$ sphere, and arrange that the fields
mapping onto $A\/$ and
$B\/$ are ``in phase'', and similarly  those mapping onto 
$B\/$ and $C\/$ are ``in
phase'', then in general the fields mapping onto 
$A\/$ and $C\/$ are {\em not} ``in phase''.
He also calculated the extent to which these last two fields are ``out of
phase'' and showed that this ``phase difference'' equals one half the
solid angle on the 
Poincar$\stackrel{\prime}{\rm e}$ sphere subtended by the spherical
triangle $ABC\/$ obtained by joining the vertices $A, B\/$ and $C\/$
by great circle arcs (geodesic arcs) on the sphere.

This fundamental and early result of Pancharatnam has found a natural
interpretation in the context of two-level quantum systems, for which
the space of pure state density matrices is again the sphere $S^2\/$.
In the modern terminology for geometric phases, one is (most often)
interested in cyclic evolution in the state space, and the
calculation of the associated geometric phase. Evolution along a
great circle arc on $S^2\/$ is particularly simple in that it can be
generated by a constant (i.e., time-independent) Hamiltonian in
such a way that the dynamical phase vanishes. A geodesic triangle on
$S^2\/$ is then the simplest and most elementary yet
 nontrivial cyclic evolution
one can imagine for a two-level system; it can be produced by a
piecewise constant Hamiltonian, and the dynamical phase can be arranged
to vanish identically. Then the meaning of Pancharatnam's result is
that the resulting geometric phase is one half the solid angle on
$S^2\/$ subtended by the triangle~\cite{geodesic-triangle}. 

We may note in passing that while Pancharatnam's original result refers 
principally to the {\em vertices} $A, B, C\/$ of a spherical triangle
on $S^2\/$, and calculates the degree of nontransitiveness of the
relation of two field amplitudes being ``in phase'', in casting it into the modern
geometric phase language we are equally concerned with the great circle
arcs connecting these vertices, since we deal with continuous cyclic
unitary or Hamiltonian evolution of pure quantum states. It is for such
evolutions that geometric phases are customarily calculated. Two more
remarks are not out of place at this point. One is that in fact geometric
phases  can be perfectly well defined for noncyclic (and even nonunitary)
evolutions, though in this paper we shall not be concerned with
them~\cite{samuel-bhandari}. The 
other is that for piecewise geodesic and overall cyclic evolutions in a
general quantum system, geometric phases get related to certain
invariants introduced by Bargmann long ago~\cite{bargmann}, 
and these invariants are in
conception very close to Pancharatnam's original motivations.

We shall refer to the ``half the solid angle'' result as the Pancharatnam
formula for geometric phases for two-level systems undergoing piecewise
geodesic cyclic evolution along a spherical triangle on $S^2\/$.
The main purposes of this paper are to: (i) present a recently developed
generalisation of the 
Poincar$\stackrel{\prime}{\rm e}$ sphere representation for pure states of
three-level quantum systems~\cite{su-3}; and (ii) to then obtain a generalisation of 
the Pancharatnam formula for such systems.

The material of this paper is arranged as follows:
Section~II outlines the generalisation of the 
Poincar$\stackrel{\prime}{\rm e}$ sphere for
three-level systems. The Poincar$\stackrel{\prime}{\rm e}$ sphere $S^2\/$ gets replaced by a
certain four-dimensional simply connected region ${\cal O}\/$ contained 
wholly within the unit sphere $S^7\/$ in eight dimensional real Euclidean
space. The transitive action of $SU(3)\/$ on ${\cal O}\/$, via eight
dimensional orthogonal rotations, some intrinsic properties of ${\cal
O}\/$, and a local coordinate system for ${\cal O} \/$, 
are described so as to assist in forming a picture of this object.
Section~III recalls very briefly the main features of the quantum
kinematic approach to the geometric phase of a general quantum system.
Both the roles of ray space geodesics and the Bargmann invariants are
highlighted. The case of two-level systems, and the statement of the
Pancharatnam formula, are then given. It is pointed out that it is a
fortunate circumstance that ray space geodesics and geodesics on $S^2\/$
happen to coincide in the case of two-level systems. Finally the general
formula for the geometric phase for any cyclic evolution of the
three-level system is given. Section~IV discusses the properties of ray
space geodesics for three-level systems, and their representation as
curves in ${\cal O}\/$. We find that the latter, while they are plane
curves, are not geodesics in the geometrical sense on $S^7\/$. We also 
find that it is possible to construct constant Hamiltonians which would
give rise to evolution along any such geodesics. Section~V puts together
the ingredients of the previous Sections to develop the generalised
Pancharatnam formula for three-level systems. This involves describing the
most general geodesic triangle for a three-level system state space, and
then computing its geometric phase. Whereas a geodesic triangle on $S^2\/$
involves three independent intrinsic parameters, for a geodesic triangle
in ${\cal O}\/$ it turns out that four independent intrinsic parameters
are needed. The geometric phase then depends on all four of these
parameters, and this is borne out by the explicit formula for the phase.
Section~VI contains some concluding remarks.    
\section{Generalisation of the 
Poincar$\stackrel{\prime}{\bf e}$ sphere representation for three-level
systems} 

We recall very briefly the salient features of the 
Poincar$\stackrel{\prime}{\rm e}$ sphere
representation for two level systems,using throughout the notations and
terminology of quantum mechanics~\cite{Poincare}.
 We deal with a two-dimensional complex
Hilbert space ${\cal H}^{(2)}\/$, unit vectors in which are denoted by
$\psi,\psi^{\prime}\cdots \/$. The density matrix corresponding to a pure
state $\psi\/$ is given by the projection $\rho= \psi\psi^{\dagger}\/$.Its
expansion in terms of the Pauli matrices $\sigma_{j}\/$ leads to the 
Poincar$\stackrel{\prime}{\rm e}$
sphere construction: 
\begin{eqnarray}
\rho&=&\psi \psi^{\dagger}=
\frac{1}{2}\left(1+{\bf n}.\mbox{\boldmath $\sigma$} \right), 
\nonumber \\
\rho^{\dagger}&=&\rho^{2} =\rho \geq 0, \mbox{Tr}\/\rho=1 \Leftrightarrow 
\nonumber \\
{\bf n}^{\star} &=&{\bf n},\, 
{\bf n}.{\bf n}=1 \Leftrightarrow {\bf n} \in S^2. 
\label{rho-s2-1}
\end{eqnarray}
Thus each pure state in  the quantum mechanical sense, or normalised ray,
corresponds in a one-to-one manner to a point on the 2-dimensional unit
sphere $S^2\/$ embedded in Euclidean three dimensional space ${\cal
R}^3\/$. Since 
\begin{eqnarray}
\rho=\psi \psi^{\dagger}=
\frac{1}{2}\left(1+{\bf n}.\mbox{\boldmath $\sigma$} \right)&,&\,
\rho^{\prime}=\psi^{\prime} \psi^{\prime \dagger}=
\frac{1}{2}\left(1+{\bf n}^{\prime}.\mbox{\boldmath $\sigma$} \right)
\Rightarrow 
\nonumber \\
\mbox{Tr}(\rho^{\prime} \rho) = \vert (\psi^{\prime}, \psi)\vert^{2}&=&
\frac{1}{2}\left(1+{\bf n}^{\prime}.{\bf n} \right)
\label{rho-s2}
\end{eqnarray}
we see that diametrically opposite points on $S^2\/$ correspond to
mutually orthogonal rays or Hilbert space vectors. Here $(\psi^{\prime},
\psi)\/$ is the inner product in ${\cal H}^{(2)}\/$. Finally, if a vector
$\psi \in {\cal H}^{(2)}\/$ is subjected to a transformation $u\in
SU(2)\/$, the 
representative point $ {\bf n} \in S^2\/$ undergoes an orthogonal 
rotation belonging to $SO(3)\/$:
\begin{eqnarray}
\psi^{\prime} &=& u \psi, \quad u \in SU(2) \Rightarrow
\nonumber \\
n_j^{\prime}&=&R_{jk}(u) n_k, 
\nonumber \\
R_{jk}(u)&=&\frac{1}{2}\mbox{Tr}(\sigma_j u \sigma_k u^{\dagger}), R(u) \in
SO(3). 
\label{n-Poincare}
\end{eqnarray}
As is well known, all elements ${\cal R} \in SO(3)\/$ are realised in
this way, and we have the coset space identifications
$S^2=SU(2)/U(1)=SO(3)/SO(2)\/$. 

Now we present the natural generalisation of this construction to
three-level systems. We deal with a 3-dimensional complex Hilbert space
${\cal H}^{(3)}\/$ , elements of which will be again denoted by $\psi,
\psi^{\prime}, \cdots\/$. The roles of $SU(2)\/$ and the Pauli matrices
$\sigma_j\/$ are now played by the group $SU(3)\/$ via its defining
representation, and the eight hermitian generators $\lambda_r\/$ in this
representation~\cite{gell-mann}:
\begin{mathletters}
\begin{eqnarray}
&&SU(3)=\left\{ A=3\times 3\, \mbox{complex matrix}\, \vert A^{\dagger}A =1,\,
\mbox{det}\, A =1\right\};
\label{su3def}
\\
\nonumber\\
&&\lambda_1=\left( \begin{array}{ccc} 
0 & 1 & 0 \\ 1 & 0 & 0 \\0 & 0 & 0 
\end{array}\right),\quad
\lambda_2=\left( \begin{array}{ccc} 
0 & -i & 0 \\ i & 0 & 0 \\0 & 0 & 0 
\end{array}\right), \quad
\lambda_3=\left( \begin{array}{ccc} 
1 & 0 & 0 \\ 0 & -1 & 0 \\0 & 0 & 0 
\end{array}\right), \quad
\lambda_4=\left( \begin{array}{ccc} 
0 & 0 & 1 \\ 0 & 0 & 0 \\1 & 0 & 0 
\end{array}\right) \nonumber \\
&&\lambda_5=\left( \begin{array}{ccc} 
0 & 0 & -i \\ 0 & 0 & 0 \\i & 0 & 0 
\end{array}\right), \, 
\lambda_6=\left( \begin{array}{ccc} 
0 & 0 & 0 \\ 0 & 0 & 1 \\0 & 1 & 0 
\end{array}\right), \, 
\lambda_7=\left( \begin{array}{ccc} 
0 & 0 & 0 \\ 0 & 0 & -i \\0 & i & 0 
\end{array}\right), \,
\lambda_8=\frac{1}{\sqrt{3}}\left( \begin{array}{ccc} 
1 & 0 & 0 \\ 0 & 1 & 0 \\0 & 0 & -2 
\end{array}\right).
\end{eqnarray} 
\end{mathletters}
The matrices $\lambda_r \/$ obey characteristic commutation and
anticommutation relations:
\begin{eqnarray}
\mbox{}&& [\lambda_r, \lambda_s]=2if_{rst}\lambda_t,  \quad
\{\lambda_r, \lambda_s \}=\frac{4}{3}\delta_{rs} + 2 d_{rst}\lambda_t;
\nonumber \\
\mbox{}&& f_{123}=1, f_{458}=f_{678}=\frac{\sqrt{3}}{2},\,
f_{147}=f_{246}=f_{257}=f_{345}=f_{516}=f_{637}= \frac{1}{2};
\nonumber \\
\mbox{}&&d_{118}=d_{228}=d_{338}=-d_{888}= \frac{1}{\sqrt{3}},\,
d_{448}=d_{558}=d_{668}=d_{778}=-\frac{1}{2\sqrt{3}},
\nonumber \\
\mbox{}&& d_{146}=d_{157}=-d_{247}=d_{256}=d_{344}=d_{355}=
-d_{366}=-d_{377}=\frac{1}{2}.
\end{eqnarray} 
Here we have given the independent nonvanishing components of the 
completely antisymmetric $f_{rst}\/$ and the completely symmetric 
$d_{rst}\/$; the former are the $SU(3)\/$ structure constants. These
$f\/$ and $d\/$ symbols allow us to define both antisymmetric and
symmetric products among real vectors ${\bf a}, {\bf b}, 
\cdots\/$ in real
eight dimensional Euclidean space ${\cal R}^8\/$, the result in each case
being another such vector~\cite{su-3}:
\begin{eqnarray}
({\bf a}_{\wedge}{\bf b})_r &=& 
f_{rst} a_s b_t, \quad {\bf a}_\wedge
{\bf b}=-{\bf b}_\wedge {\bf a}; \nonumber \\ 
({\bf a} \star {\bf b})_r &=& \sqrt{3}d_{rst} a_s b_t, 
\quad {\bf a} \star
{\bf b}={\bf b} \star {\bf a}.
\label{products}
\end{eqnarray} 
The significance of these definitions is that 
${\bf a}\wedge {\bf b}\/$ and 
${\bf a} \star {\bf b} \/$ transform just as
${\bf a}\/$ and ${\bf b}\/$ do, under
the eight dimensional adjoint representation of $SU(3)\/$.
The matrices of this representation are defined similarly to  
eq.(~\ref{n-Poincare}):
\begin{eqnarray}
A \in SU(3) \rightarrow D_{rs}(A) &=& \frac{1}{2} \mbox{Tr}(\lambda_r A
\lambda_s  A^{\dagger}), \nonumber \\
D(A^{\prime})  D(A) &=& D(A^{\prime} A), \nonumber \\
D(A) &\in& SO(8).
\end{eqnarray} 
However in contrast to the $SU(2)-SO(3)\/$ case, here the matrices 
$D(A)\/$ that arise are only an eight-parameter family, and so a very
small portion indeed of the full twenty-eight-parameter group $SO(8)\/$. 
In any case the required properties of the products~(\ref{products}) are:
\begin{eqnarray}
D(A) {\bf a}_{\wedge} D(A) 
{\bf b} &=& D(A) ({\bf a}_{\wedge}{\bf b}),
\nonumber \\
D(A) {\bf a} \star D(A) 
{\bf b} &=& D(A) ({\bf a} \star {\bf b}).
\end{eqnarray}

With this background, we can handle general pure state density 
matrices for three level systems~\cite{su-3}. Given a normalised $\psi \in {\cal
H}^{(3)}\/$, we form the density matrix $\rho=\psi \psi^{\dagger}\/$
and expand it in terms of the unit matrix and the $\lambda_{r}\/$:
\begin{eqnarray}
\psi \in {\cal H}^3 &,& \quad (\psi, \psi)=1:
\nonumber \\
\rho&=&\psi \psi^{\dagger} = 
\frac{1}{3}(1+\sqrt{3} {\bf n}.{\mbox{\boldmath $\lambda$}} ).
\label{def-rho}
\end{eqnarray}
We then find in place of eq~(\ref{rho-s2}):
\begin{equation}
\rho^{\dagger}=\rho^{2} =\rho \geq 0, \mbox{Tr}\/\rho=1 \Leftrightarrow 
{\bf n}^{\star} ={\bf n},\, 
{\bf n}.{\bf n}=1, \, 
{\bf n}\star {\bf n}={\bf n} 
\end{equation}
Thus each normalised ray for the three level system  corresponds
uniquely in a one-to-one manner to a unit vector ${\bf n} \in S^7\/$,
the seven dimensional unit sphere in ${\cal R}^8\/$, which moreover obeys
the condition ${\bf n}\star {\bf n} ={\bf n}\/$. 
The set of all such real unit
vectors in ${\cal R}^8\/$, a subset of $S^7\/$, is the analogue of the
Poincar$\stackrel{\prime}{\rm e}$ sphere for three-level systems. Since it is in fact a very small
part of $S^7\/$, we give it a special symbol:
\begin{equation}
{\cal O} =\left\{ {\bf n} 
\in {\cal R}^8 \vert{\bf n}. {\bf n} =1, 
\quad 
{\bf n}\star {\bf n} ={\bf n}  
\right\} \subset S^7 \subset {\cal R}^8.
\label{O-def}
\end{equation} 
This set ${\cal O}\/$ is a connected, simply connected four dimensional
region contained in $S^7\/$, and its points correspond one-to-one to pure
states of a three-level system. It is in fact a representation of the
coset space $SU(3)/U(2)\/$

Some interesting geometric properties of ${\cal O}\/$ may be mentioned.
For two unit vectors $\psi,\, \psi^{\prime} \in {\cal H}^{(3)}\/$, we find: 
\begin{eqnarray}
\rho=\psi \psi^{\dagger}, \, \rho^{\prime}=\psi^{\prime}
\psi^{\prime \dagger} & \Rightarrow & \mbox{Tr}(\rho^{\prime}\rho)=
\vert (\psi^{\prime}, \psi)\vert^{2} = \frac{1}{3} (1+2
{\bf n}^{\prime}.{\bf n}),
\nonumber \\
0 \leq \mbox{Tr}(\rho^{\prime} \rho) \leq 1 & \Leftrightarrow & 0 \leq
\cos^{-1}({\bf n}^{\prime}.{\bf n}) \leq \frac{2 \pi}{3} .
\end{eqnarray}  
Thus mutually orthogonal vectors in ${\cal H}^{(3)}\/$ do not lead to
antipodal or diametrically opposite points on ${\cal O}\/$, but rather to
points with a  maximum opening angle of $\frac{2 \pi}{3}\/$ radians. Indeed,
if ${\bf n} \in {\cal O}\/$, then $-{\bf n} \not\in {\cal O}\/$. If one takes
the three canonical basis vectors of ${\cal H}^{(3)}\/$ as usual, they
lead to three distinguished points or ``poles'' on ${\cal O}\/$:
\begin{eqnarray}
\left( 1,0,0 \right)^{T} \rightarrow 
n_3=\frac{\sqrt{3}}{2},\, n_8=\frac{1}{2}, \, \mbox{rest zero};
\nonumber \\
\left( 0,1,0 \right)^{T} \rightarrow 
n_3=-\frac{\sqrt{3}}{2},\, n_8=\frac{1}{2}, \, \mbox{rest zero};
\nonumber \\
\left( 0,0,1 \right)^{T} \rightarrow 
n_8=-1, \, \mbox{rest zero},
\end{eqnarray}
each making an angle of $\frac{2\pi}{3}\/$ with any other. These properties of
${\cal O }\/$  may help one make some sort of mental picture of this
geometrical object embedded in $S^{7}\/$.

The action of $SU(3)\/$ on vectors in ${\cal H}^{(3)}\/$ leads to adjoint
action on ${\cal O}\/$:
\begin{equation}
A\in SU(3) \quad :\quad \psi^{\prime}=
A\psi\Rightarrow {\bf n}^{\prime}=D(A){\bf n}.
\end{equation}
Thus one has here a (small set of) rigid eight-dimensional orthogonal
rotations, which will prove convenient later on. Moreover, since ${\cal
O}\/$ is the coset space $SU(3)/U(2)\/$, this adjoint action of $SU(3)\/$
on ${\cal O}\/$ is transitive. This will also be exploited later. General
$SO(8)\/$ rotations of course do not preserve the region $ {\cal O}\/$ of
$S^7\/$.

For practical calculations it is convenient to introduce four independent
local angle type variables which can be used as coordinates over (almost
all of) ${\cal O}\/$. Let us write a general unit vetor $\psi \in {\cal
H}^{(3)}\/$ as:
\begin{equation}
\psi = \left( \begin{array}{c} \psi_1\\ \psi_2 \\ \psi_3 \end{array}
\right) ,\, \psi^{\dagger}\psi=\vert \psi_1 \vert^{2} + 
\vert \psi_2 \vert^{2} + \vert \psi_3 \vert^{2}=1.
\end{equation}
Then omitting the part of ${\cal O}\/$ corresponding to $\psi_3=0\/$
( this is a two-dimensional region, essentially an $S^2\/$, see below), 
over the
rest of ${\cal O}\/$ we introduce $\theta, \phi, \chi_1, \chi_2\/$ in
this way~\cite{local-var}:
\begin{eqnarray}
&\left(\psi_1, \psi_2,  \psi_3 \right) = (\mbox{overall phase})
\left(e^{i \chi_1}\sin{\theta}\cos{\phi},
 e^{i \chi_2}\sin{\theta}\sin{\phi}, \cos{\theta} \right) &
\nonumber \\
&0 \leq \theta < \frac{\pi}{2}, \, 0 \leq \phi \leq \frac{\pi}{2},\,
0 \leq \chi_1, \chi_2 <2 \pi.&
\label{local-co-O}
\end{eqnarray}
The limits on $\theta, \phi \/$ reflect the nonvanishing of $\psi_3\/$,
and the fact that the real three-dimensional  unit vector 
$(\vert \psi_1 \vert, \vert \psi_2 \vert , \vert \psi_3 \vert)\/$ has
non-negative components. Thus $\theta, \phi\/$ denotes a point on the
first octant of an $S^2\/$. Given $\vert \psi_3\vert >0\/$, $\chi_1\/$ is
the phase of $\psi_1\/$ relative to $\psi_3\/$ ( and is well defined
except when $\phi=\frac{\pi}{2})\/$; and $\chi_2\/$ is
the phase of $\psi_2\/$ relative to $\psi_3\/$ ( and is well defined
except when $\phi=0$). All this is shown in Fig.1. We need to remember
that $\theta, \phi\/$ determine the magnitudes of the components of
$\psi\/$, while $\chi_1, \chi_2\/$ give their relative phases. All four
taken together determine  one point in the portion of ${\cal O}\/$
with $\psi_3 \not=0\/$. 
\input epsf.tex
\begin{figure} 
\hspace{3.5cm}\epsfxsize=10cm \epsfbox{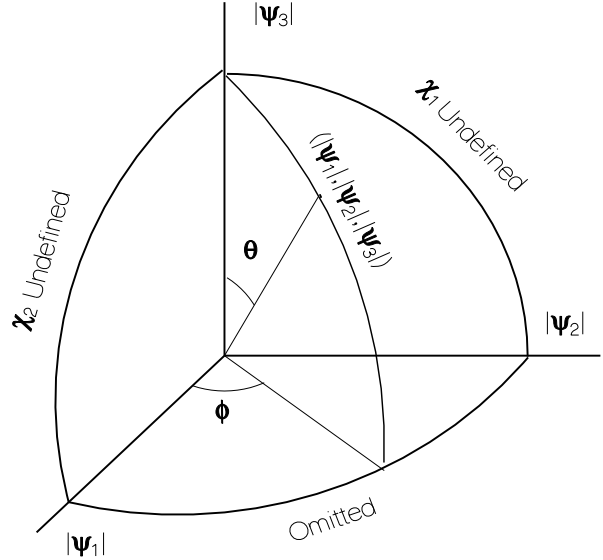}
\caption{A pictorial description of almost all of the space 
${\cal O}\/$ (the space of all states of the three-level system).
The local coordinates $\theta, \phi, \chi_1, \chi_2\/$ are such that
$\theta\/$ and $\phi \/$ define a point 
in the positive octant of $S^2\/$ i.e. $ 0 \leq
\theta, \phi \leq \frac{\textstyle \pi}{\textstyle 2}\/$ and  for a given
point on this octant we have a torus defined by two angle variable $0
\leq  \chi_1, \chi_2 < 2 \pi \/$.}
\end{figure}

We can easily obtain the expressions for $n_r\/$ in these
local coordinates.
Combining  eqs.~(\ref{def-rho},~\ref{local-co-O}) we get:
\begin{eqnarray}
n_r&=&\frac{\sqrt{3}}{2} \psi^{\dagger} \lambda_r \psi:
\nonumber \\
{\bf n}&=& \sqrt{3}\left(\frac{\mbox{}}{\mbox{}}
\sin^2{\theta} \sin{\phi}\cos{\phi}\cos{(\chi_2-\chi_1)}, \,
\sin^2{\theta} \sin{\phi}\cos{\phi}\sin{(\chi_2-\chi_1)}, \right.
\nonumber \\
&&\quad\quad\quad\frac{1}{2} \sin^2{\theta}(\cos^2{\phi}-\sin^2{\phi}), \,
\sin{\theta}\cos{\theta} \cos{\phi} \cos{\chi_1}, 
\nonumber \\
&&\quad\quad\quad-\sin{\theta}\cos{\theta} \cos{\phi} \sin{\chi_1}, \,
\sin{\theta}\cos{\theta} \sin{\phi} \cos{\chi_2},
\nonumber \\
&&\quad\quad\quad\left. -\sin{\theta}\cos{\theta} \sin{\phi} \sin{\chi_2},\,
\frac{1}{2\sqrt{3}}(1-3\cos^2\theta) \right). 
\end{eqnarray}

The recovery of the Poincar$\stackrel{\prime}{\rm e}$ sphere $S^2\/$ for a two-dimensional
subspace of ${\cal H}^{(3)}\/$ is straightforward. Consider as an example
vectors $\psi \in {\cal H}^{(3)}\/$ with vanishing third component (just
the points of ${\cal O}\/$ omitted in the local
coordinatisation~(\ref{local-co-O}) of ${\cal O}\/$):
\begin{eqnarray}
\psi=\left(\begin{array}{c}\psi_1\\ \psi_2 \\ 0 \end{array}\right)=
\left(\begin{array}{c} e^{i\chi_1}\, \cos{\phi}\\ e^{i \chi_2}\,\sin{\phi}\\0
 \end{array}\right),
\nonumber \\
0\leq\phi\leq\frac{\pi}{2}, 0 \leq \chi_1\/,\, \chi_2 < 2 \pi
\label{s2-recover}
\end{eqnarray}
Then the eight-vector ${\bf n}\/$ has only four nonvanishing components:
\begin{eqnarray}
\left(n_1,\,n_2,\, n_3\right)&=&\frac{\sqrt{3}}{2}\/\left( \sin{2\phi}
\cos{(\chi_2-\chi_1)},\/ \sin{2 \phi} \sin{(\chi_2-\chi_1)},\/ \cos{2
\phi} \right),
\nonumber \\
n_8&=&\frac{1}{2},
\nonumber \\
n_4&=&n_5=n_6=n_7=0.
\end{eqnarray}
As $\phi\/$ and $(\chi_2-\chi_1)\/$ vary in the appropriate ranges, we
see that we obtain a certain sphere $S^2\/$ embedded within ${\cal O}\/$,
centred on the point $(0,0,0,0,0,0,0,\frac{\textstyle 1}{\textstyle 2})\/$,
of radius $\frac{\textstyle \sqrt{3}}{\textstyle 2}\/$, and contained
entirely within the $1-2-3-8\/$ subspace of ${\cal R}^{8}\/$. If we
consider two dimensional subspaces in ${\cal H}^{(3)}\/$ different 
from~(\ref{s2-recover}), we clearly obtain $SU(3)\/$ transforms of the above
situation. All these various $S^2$'s are off-centre in ${\cal
R}^{8}\/$: indeed their centres lie on a sphere in ${\cal R}^8\/$
centred at the origin of ${\cal R}^8\/$ and of radius 
$\frac{\textstyle 1}{\textstyle 2}\/$
\section{Background to the geometric phase and Pancharatnam's formula}
Consider a general quantum mechanical system whose pure states are
described by unit vectors in a complex Hilbert space ${\cal H}\/$ of any
dimension. The corresponding ray space will be denoted by ${\cal R}\/$. Let
${\cal C}\/$ be a continuous piecewise smooth parametrised curve of unit
vectors in ${\cal H}\/$:
\begin{equation}
{\cal C} = \left\{ \psi(s) \vert s_1 \leq s \leq s_2 \right \} \subset
{\cal H},
\end{equation}
and let $C\/$ be its image in ${\cal R}\/$, likewise continuous and
piecewise smooth:
\begin{equation}
 C = \left\{ \rho(s) =
\psi(s)\psi(s)^{\dagger} \vert s_1 \leq s \leq s_2 \right \} \subset
{\cal R}.
\end{equation}
In case $\psi(s_2)\/$ and $\psi(s_1)\/$ determine the same ray, and in
particular $\psi(s_2) = \psi(s_1)\/$ in which case ${\cal C}\/$ is
closed, the image $C\/$ is closed; however in general we need not assume
this. The geometric phase associated with $C\/$ is the difference between
a total (or Pancharatnam) phase and a dynamical phase, each of which
is a functional of ${\cal C}\/$~\cite{simon-mukunda-1}:
\begin{eqnarray}
\varphi_g[C] & = & \varphi_p[{\cal C}]-\varphi_{\rm dyn}[{\cal C}],
\nonumber \\
\varphi_p[{\cal C}]&=&{\rm arg}(\psi(s_1),\psi(s_2)),
\nonumber \\
\varphi_{\rm dyn}[{\cal C}]&=& {\rm Im}\int\limits_{s_1}^{s_2}ds
(\psi(s),\dot{\psi}(s)).
\end{eqnarray}
The quantity $\varphi_g[C]\/$ is invariant under both local smooth phase
changes in $\psi(s)\/$, and under smooth reparametrisations - for these
reasons it is a geometric quantity dependent on $C\/$ rather than on 
${\cal C}\/$.

In this context an important role is played by {\em geodesics in the space
${\cal R}$}~\cite{simon-mukunda-1}. Given the continuous curve $C \subset {\cal R}\/$( with
nonorthogonal end points for definiteness), 
a nondegenerate positive definite length functional ${\cal L}[C]\/$ 
can be set up, which is also
reparametrisation invariant:
\begin{equation}
{\cal L}[C]=\int\limits^{s_2}_{s_1} ds \left\{
 (\dot{\psi}(s),\dot{\psi}(s))-
(\psi(s),\dot{\psi}(s))(\dot{\psi}(s),\psi(s)) \right\}^{\frac{1}{2}}.
\end{equation}
 Extremising this functional ( with fixed end
points), we arrive at the concept of geodesics in ${\cal R}\/$. Any
Hilbert space lift of such a geodesic, with any choice of parametrisation,
may then be called a geodesic in ${\cal H}\/$. It then turns out that
every geodesic in ${\cal R}\/$ has vanishing geometric phase:
\begin{equation}
C=\mbox{geodesic in ${\cal R}$} \Rightarrow \varphi_g[C]=0,
\label{geo-def}
\end{equation}
and this accounts for their importance. The simplest description of a
geodesic, which can always be achieved, is as follows.
Let the end points of $C\/$ be $\rho^{(1)}\/$ and $\rho^{(2)}\/$, assumed
nonorthogonal, and choose unit vectors $\psi^{(1)}, \psi^{(2)}\/$ such
that:
\begin{eqnarray}
\rho^{(1)} =\psi^{(1)}\psi^{(1) \dagger}&,& \quad
\rho^{(2)} =\psi^{(2)}\psi^{(2) \dagger}
\nonumber \\
(\psi^{(1)}, \psi^{(2)}) &=& \mbox{real positive}.
\end{eqnarray}
Thus $\psi^{(1)}\/$ and $\psi^{(2)}\/$ are ``in phase'' in the
Pancharatnam sense. Then the geodesic $C_{\rm geo} \subset {\cal R}\/$
connecting $\rho^{(1)}\/$ to $\rho^{(2)}\/$ is the ray space image of
the following curve ${\cal C}_{\rm geo} \subset {\cal H}\/$:
\begin{eqnarray}
{\cal C}_{\rm geo} &=& 
\left\{ \psi(s) \vert 0 \leq s \leq s_0 \right \},
\nonumber \\
\psi(s)&=&\psi(0)\cos{s} + \dot{\psi}(0)\sin{s},
\nonumber \\
\psi(0)&=& \psi^{(1)}, \quad \dot{\psi}(0)=
\left(\psi^{(2)}-\psi^{(1)}(\psi^{(1)}, \psi^{(2)})\right)/
\left(1-(\psi^{(1)},\psi^{(2)})^2 \right)^{\frac{1}{2}},
\nonumber \\
s_0 &=& \cos^{-1}( \psi^{(1)},\psi^{(2)}).
\label{geo-hilbert}
\end{eqnarray} 

Exploiting the fundamental result~(\ref{geo-def}) we obtain a very
attractive expression for the geometric phase in the following particular
situation. Choose a set of points $\rho^{(1)}, \rho^{(2)},\cdots
,\rho^{(n)} \in {\cal R}\/$ in a definite sequence, assume for definiteness
that no two consecutive points are mutually orthogonal, and also that 
$\rho^{(n)}\/$ and $\rho^{(1)}\/$ are nonorthogonal. Connect
$\rho^{(1)}\/$ to $\rho^{(2)}\/$, $\rho^{(2)}\/$ to
$\rho^{(3)},\/\cdots\/,\rho^{(n)}\/$ to $\rho^{(1)}\/$ by geodesic arcs,
so that we obtain a closed curve $C \subset {\cal R}\/$ in the form of 
an $n$-sided polygon made up of geodesic pieces. Then we have~\cite{simon-mukunda-1}:
\begin{eqnarray}
C &= & \mbox{geodesic polygon in ${\cal R}\/$ with vertices $\rho^{(1)},
\rho^{(2)}, \cdots,\rho^{(n)}$}:
\nonumber \\
\varphi_g[C] & = &-{\rm arg}(\psi^{(1)},\psi^{(2)})(\psi^{(2)},\psi^{(3)})
\cdots (\psi^{(n)},\psi^{(1)})
\nonumber\\
&=&-{\rm arg}\, {\rm Tr}(\rho^{(1)}\rho^{(2)} \cdots \rho^{(n)}),
\nonumber \\
\rho^{(1)}&=&\psi^{(1)}\psi^{(1) \dagger},\,
\rho^{(2)}=\psi^{(2)}\psi^{(2) \dagger},\cdots,
\rho^{(n)}=\psi^{(n)}\psi^{(n) \dagger}.
\label{bargmann}
\end{eqnarray}
Here it is evident that the phases of the vectors $\psi^{(1)},\psi^{(2)},
\cdots, \psi^{(n)}\/$ can be freely chosen. The result~(\ref{bargmann})
connects the geometric phase for a closed polygon to the Bargmann
invariant of quantum mechanics, the expression 
$(\psi^{(1)},\psi^{(2)})
(\psi^{(2)},\psi^{(3)}) \cdots (\psi^{(n)},\psi^{(1)})\/$, namely: the
former is the negative of the argument of the latter. The point to
emphasize is that the definition of the Bargmann invariant requires 
specifying just the vertices of the polygon, while the definition of the
geometric phase requires also connecting them in sequence by geodesic
arcs so that we have a closed loop $C \subset {\cal R}$.

With this background from the general theory of the geometric phase we
relate these results to the case of two level systems, quote the
Pancharatnam formula and then give the general expression for
$\varphi_g[C]\/$ for three-level systems. For two-level systems we
have seen that the space ${\cal R}\/$ is the Poincar$\stackrel{\prime}{\rm e}$ sphere $S^2\/$.
It is now a happy coincidence that geodesics in ${\cal R}\/$ map
exactly on to geodesics on $S^2\/$ in the more familiar Euclidean
sense. This can be seen as follows.
Without loss of generality, by using a suitable $SU(2)\/$
transformation, we may assume that the points $\rho^{(1)}, \rho^{(2)}\/$ to be
connected by a geodesic are the points ${\bf n}^{(1)}=(0,\, 0,\, 1), {\bf
n}^{(2)}=(\sin 2\alpha , \,0 , \,\cos 2\alpha)\/$ on $S^2\/$, with
representative vectors $\psi^{(1)}, \psi^{(2)} \in {\cal
H}^{(2)}\/$ chosen as follows:
\begin{equation}
\psi^{(1)}=\left(\begin{array}{c} 1\\0 \end{array}\right)\,\quad
\psi^{(2)}=\left(\begin{array}{c} \cos{\alpha} \\ \sin{\alpha}
\label{s2-vectors}
\end{array}\right).
\end{equation}
Then applying the result~(\ref{geo-hilbert}), the ray space geodesic
connecting $\rho^{(1)}\/$ to $\rho^{(2)}\/$ is determined  as follows:
\begin{eqnarray}
\psi(s)&=&\left(\begin{array}{c} \cos{s} \\ \sin{s} \end{array}\right);
\nonumber \\
{\bf n}(s)&=&{\rm Tr} \rho(s) \mbox{\boldmath $\sigma$} =
(\psi(s),\mbox{\boldmath $\sigma$}\psi(s))
\nonumber \\
&=&\left(\sin{2 s}\, ,0 \, , \cos{2 s}\right)\,, \quad 0\leq s \leq 
\alpha.
\label{s2-geodesic}
\end{eqnarray}
We see that the curve described by ${\bf n}(s)\/$ on $S^2\/$ is indeed
a great circle arc, a part of the ``Greenwich meridian''; and by the
action of $SU(2)\/$ on ${\cal R}\/$ translated into the action of
$SO(3)\/$ on $S^2\/$, we conclude that {\em any} ray space geodesic
in ${\cal R}\/$ appears as some great circle arc on $S^2\/$. Thus the
two definitions of geodesics do coincide in this case.

Let now $A, B,\/$ and $C\/$ be any three points on $S^2\/$, no two
being diametrically opposite to one another. Joining them by great
circle arcs (each less than $\pi$ in extent) we get a geodesic
triangle $\triangle(A, B, C)\/$ on $S^2\/$. Then the Pancharatnam
formula~\cite{pancharatnam}~\cite{bary-2}~\cite{rajaram} is the statement that for any two level system:
\begin{eqnarray}
\mbox{}&\varphi_g[\triangle(A, B, C)]=\frac{1}{2}\Omega,&
\nonumber \\
\mbox{}&\Omega=\mbox{solid angle subtended by the triangle $A,B,C\/$ at the
origin of $S^2$\/}&.
\label{panch-solid}
\end{eqnarray}
Here the right hand side is interpreted to be positive(negative) if, as
viewed from outside of $S^2\/$, the triangle $ABC\/$ is described in the
counter clockwise (clockwise) sense.
It is this formula that we shall generalise in Section~V. 

Now we give the general formula for geometric phases for  three
level systems~\cite{su-3}. Consider a closed loop $C\subset {\cal O}\/$, and
assume for definiteness that $\psi_3 \neq 0\/$ throughout. Then
$\varphi_g[C]\/$ is given by the following integral along $C\/$:  
\begin{equation}
\varphi_g[C]=-\ointop\limits_{C\subset {\cal O}}
\sin^{2}{\theta}(\cos^{2}{\phi}\/ d \chi_1 + \sin^2{\phi}\/ d \chi_2)
\end{equation}
We note that this formula holds when $C\/$ is a closed loop. In the
succeeding Sections we develop the properties of geodesics in ${\cal
O}\/$ and then generalise eq.~(\ref{panch-solid})

\section{Ray space geodesics for three-level systems}
We have seen that the transitive action of $SU(3)\/$ on the ray space
${\cal O}\/$ for three-level systems is given by (an eight-parameter
subset of) rigid orthogonal
rotations in real Euclidean eight dimensional space, when points of
${\cal O}\/$ are identified with vectors ${\bf n}\/$ as in 
eq~(\ref{O-def}). Based on this we may describe the details of
any one conveniently chosen geodesic in ${\cal O}\/$; and then any
other would be a suitable $SO(8)\/$ transform of this one, so that
the geometrical shape and structure in an intrinsic sense are
unaltered.

Guided by the constructions of Section~III, 
eq.~(\ref{s2-vectors},~\ref{s2-geodesic}) let us choose two points
in ${\cal O}\/$ correponding to the following two unit vectors in
${\cal H}^{(3)}\/$:
\begin{eqnarray}
\psi^{(1)}&=&\left(\begin{array}{c} 0\\0\\1 \end{array}\right)\quad\,\quad
\psi^{(2)}=\left(\begin{array}{c} 0\\ \sin{\alpha} \\ \cos{\alpha}
\end{array}\right)
\nonumber \\
{\bf n}^{(1)}&=&\left(0 \,,\,0\,,\,0\,,\,0\,,\,0\,,\,0\,,\,0\,,\, -1\right)
\nonumber\\
{\bf n}^{(2)}&=&\frac{\sqrt{3}}{2}\left(0 \,,\,0\,,\,-\sin^2{\alpha}\,,\,0\,,\,0\,
,\,2 \sin{\alpha} \cos{\alpha}\,,\,0\,,\, \frac{1}{\sqrt{3}}(1-3
\cos^2{\alpha})  \right)
\label{vectors-s7}
\end{eqnarray}
It is easy to see that, given {\em any} pair of three level pure
state density matrices $\rho^{(1)}\/$ and $\rho^{(2)}\/$ such that 
Tr$(\rho^{(1)}\rho^{(2)})=\cos^2{\alpha} > 0\/$, we can exploit the
action of $SU(3)\/$ on ${\cal H}^{(3)}\/$, and freedom of phases, to put 
$\rho^{(1)}\/$ and $\rho^{(2)}\/$ into the configurations
corresponding to the vectors $\psi^{(1)}$ and $\psi^{(2)}$   above.
Then the ray space geodesic $C^{(0)}_{\rm geo}\/$ connecting ${\bf
n}^{(1)}\/$ and ${\bf n}^{(2)}$ is easily found on the basis of
the general formula~(\ref{geo-hilbert}):  
\begin{eqnarray}
\mbox{}&&\psi(s)=\left(\begin{array}{c} 0\\ \sin{s} \\ \cos{s}
\end{array}\right);
\nonumber \\
C^{(0)}_{\rm geo}:\quad {\bf n}(s)&=&\frac{\sqrt{3}}{2} \psi(s)^{\dagger} {\bf
\lambda} \psi(s)
\nonumber \\
=\frac{\sqrt{3}}{2}\left(0 \,,\,0\,,\,-\sin^2{s}\,\right.,& 0 &,\left.\,0\,
,\,2 \sin{s} \cos{s}\,,\,0\,,\, \frac{1}{\sqrt{3}}(1-3
\cos^2{s})  \right),\, 0\leq s \leq \alpha.
\label{three-geo}
\end{eqnarray}
This curve $\{ {\bf n}(s)\} \subset {\cal O}\/$ has only three
nonvanishing components, namely, $n_3(s), n_6(s)\/$ and $n_8(s)\/$.
The interesting questions are whether it is a plane curve, whether it
coincides with a geodesic arc as defined in the sense of eight
dimensional Euclidean geometry on $S^7\/$, and whether it has any
other important geometric features.

The first observation we may make is that since the components 
$n_1, n_2, n_4, n_5, n_7\/$ all vanish, this curve $C^{(0)}_{\rm geo}\/$
lies entirely in a three dimensional subspace of ${\cal R}^8\/$. By
$SU(3)\/$ action this statement is then true for all ray space
geodesics $C_{\rm geo}\/$ when drawn in ${\cal O}\/$. Next we remark
that a geodesic on $S^7\/$ in the Euclidean sense would be part of
the intersection of a two dimensional plane in ${\cal R}^8\/${\em passing
through the origin}, with $S^7\/$. We can immediately see that
$C^{(0)}_{\rm geo}\/$ {\em is not of this kind}; for example, the three
vectors ${\bf n}(0), {\bf n}(\frac{\textstyle \alpha}{\textstyle 2}) ,{\bf
n}(\alpha)\/$ are easily checked to be linearly
independent. Thus in striking contrast to the situation on the
Poincar$\stackrel{\prime}{\rm e}$ sphere $S^2\/$ for two level systems, here {\em ray space
geodesics on ${\cal O}\/$ are not geodesic arcs in the sense of
Euclidean eight dimensional geometry}.     

Nevertheless we can show that $C^{(0)}_{\rm geo}\/$ is a plane curve,
though the plane on which it lies is off center in ${\cal R}^8\/$,
i.e. it does not pass through the origin or the center of $S^7\/$.
For this we need to work with the combinations 
$\frac{\textstyle \sqrt{3} n_3 + n_8}{\textstyle 2}, 
\frac{\textstyle n_3 - \sqrt{3} n_8}{\textstyle 2}\/$
obtained from $n_3, n_8\/$ by an orthogonal transformation. Then we find:
\begin{eqnarray}
\frac{\sqrt{3} n_3(s) + n_8(s)}{2}&=&-\frac{1}{2}\,, \nonumber \\
 n_6(s)&=&\sqrt{3} \sin{s} \cos{s}\,, \nonumber \\
\frac{ n_3(s) - \sqrt{3} n_8(s)}{2}&=&
-\frac{\sqrt{3}}{2}(1-2 \cos^2{s}) \nonumber \\
\end{eqnarray}
Thus, in the three dimensional $n_3-n_6-n_8\/$ subspace, this geodesic
$C^{(0)}_{\rm geo}$ is a curve lying in the two dimensional plane 
$\frac{\textstyle \sqrt{3} n_3 + n_8}{\textstyle 2}=-\frac{1}{2}\/$. 
Ray space geodesics
$C_{\rm geo}\/$ for three level systems, when described as curves
$\{ {\bf n}(s)\} \subset {\cal O}\/$, are really like arcs of constant
latitude on $S^2\/$, and not geodesic arcs at all. For this reason,
hereafter the term geodesic will refer exclusively to ray space
geodesics, with no further qualification.

In the description of ${\cal O}\/$ using coordinates $\theta , \phi
,\chi_1, \chi_2\/$ the geodesic $C^{(0)}_{\rm geo}\/$ appears as follows:
\begin{equation}
C^{(0)}_{\rm geo} \quad : \theta(s)=s ,\,
\phi(s)=\frac{\pi}{2} , \,\chi_1(s) \mbox{undefined} ,\,
\chi_2(s)=0,\,0\leq s \leq \alpha.
\end{equation}
Thus the projection of $C^{(0)}_{\rm geo}\/$ onto the octant of $S^2\/$
with spherical polar angles $\theta, \phi\/$ happens to be a great
circle arc in the usual sense. However this is not expected to be
true for a general geodesic $C_{\rm geo} \subset {\cal O}\/$.

One can now ask for the most general three level system Hamiltonian
which reproduces the evolution of $\psi(s)\/$ with respect to $s\/$,
as already specified in eq.~(\ref{three-geo}), and whether it can
be independent of $s\/$. A general hermitian Hamiltonian can be
written in terms of nine real functions of $s\/$ as:
\begin{equation}
H(s) = h_0(s) + {\bf h}(s).\mbox{\boldmath $\lambda$}, 
\end{equation}
while the quantum mechanical evolution equations for $\psi(s)\/$ and 
$\rho(s)\/$ become:
\begin{eqnarray}
i\/\frac{d}{ds}\psi(s) &=& H(s) \psi(s)\,,
\nonumber \\
\frac{d}{ds}{\bf n}(s) &=& 2 {\bf h}(s)_{\wedge} {\bf n}(s).
\end{eqnarray}
An easy calculation shows that the most general $H(s)\/$ has four
independent real arbitrary functions of $s\/$, namely $a(s),\,b(s),\,
c(s),\, d(s)\/$ in:
\begin{eqnarray}
{\bf h}(s)&=&\left( a(s) \cos{s} , b(s) \cos{s} , \sqrt{3}\/ c(s) +
d(s)\/(\cos^2{s}-\sin^2{s}), \right.
\nonumber \\
\mbox{}&&\quad\quad \left.-a(s)\/ \sin{s},\/  -b(s)\/\sin{s}, 
d(s)\/\cos{s}\sin{s}, -1, c(s)^{\mbox{}} \right)
\nonumber\\
h_0(s)&=&\frac{2}{\sqrt{3}}\/c(s)-d(s)\/\sin^2{s}.
\end{eqnarray}
Moreover we also find with this general Hamiltonian :
\begin{eqnarray}
{\rm Tr}(\rho(s) H(s))&=&(\psi(s),H(s)\psi(s))
\nonumber \\
&=&h_0(s)+\frac{2}{\sqrt{3}}{\bf n}(s).{\bf h}(s)
\nonumber\\
&=&0.
\end{eqnarray}
This is consistent with the vanishing of the 
dynamical phase as the state evolves along $\{\psi(s)\}\/$ of
eq~(\ref{three-geo}), which is directly checked to be true.
 If we in addition wish to have a constant Hamiltonian 
producing this evolution, we must set $a(s)
=b(s)=d(s)=0, c(s)=c_0=\mbox{constant}\/$ and then we get:
\begin{equation}
H^{(0)}_{\rm const.}=\left( \frac{2}{\sqrt{3}} {\bf 1}+\sqrt{3}\lambda_3+
\lambda_8
\right) c_0 -\lambda_7.
\end{equation}
It is possible to express this simple constant Hamiltonian
directly in terms of the end points ${\bf n}^{(1)},{\bf n}^{(2)} \/$
of the geodesic $C^{(0)}_{\rm geo}$ given in eq.~(\ref{vectors-s7}). Namely one
finds: 
\begin{equation}
({\bf n}^{(1)}\mbox{}_\wedge{\bf n}^{(2)})_r=
-\frac{3}{2}\left(\sin{\alpha} \cos{\alpha}\right)\delta_{r 7}
\end{equation}
which, if we set the constant $c_0\/$ to zero,  leads to: 
\begin{equation}
H^{(0)}_{const.}=
 \frac{{\bf n}^{(1)}\mbox{}_\wedge{\bf n}^{(2)}.\mbox{\boldmath $\lambda$}}
{\vert{\bf n}^{(1)}\mbox{}_\wedge{\bf n}^{(2)}\vert}\,.
\end{equation}

We can now generalise these results to {\rm any} geodesic $C_{\rm
geo} \subset {\cal O}\/$ connecting any two points 
${\bf n}, {\bf n}^{\prime} \in {\cal O}\/$.
Writing ${\bf n}.{\bf n}^{\prime}=\frac{1}{2}(3 \cos^2{\alpha}-1)\/$, 
we have the result that the constant Hamiltonian
\begin{equation}
H_{const.}=
\frac{2\,{\bf n}\mbox{}_\wedge{\bf n}^{\prime}.\mbox{\boldmath $\lambda$}}
{3\/ \sin{\alpha} \cos{\alpha}},
\end{equation}
can produce evolution along the geodesic $C_{\rm geo}\/$.
This generalises well known results in the case of two-level systems~\cite{geodesic-triangle}.
\section{The Pancharatnam geometric 
phase formula for three-level systems}
The purpose of this Section is to obtain the generalisation of the
result~(\ref{panch-solid}) for three-level systems. 
For this we must determine the
geometrical description of the most general geodesic 
triangle on ${\cal O}\/$, upto
an overall $SU(3)\/$ transformation, and then use the general
connection~(\ref{bargmann}) to find the geometric phase associated with
this triangle.

Let $A, B\/$ and $C\/$ be three general points in ${\cal O}\/$ such
that no two of them enclose an angle of $\frac{\textstyle 2
\pi}{\textstyle 3}\/$ radians.
Using the freedom of common $SU(3)\/$ action we can transport $A\/$ to a 
position $A^{(0)}\/$ with a representative  vector of the form
$\psi^{(1)}\/$ of eq~(\ref{vectors-s7}). In this process let $B\/$ and $C\/$
move to locations $B^{\prime}\/$ and $C^{\prime}\/$:

\begin{eqnarray}
A,B,C \shortstack[l]{{$SU(3)$}\\ {$\longrightarrow$}} 
A^{(0)},B^{\prime}, C^{\prime}\quad :  
\nonumber \\
A^{(0)}\rightarrow \rho^{(1)}=\psi^{(1)}\psi^{(1) \dagger}
\quad , \quad \psi^{(1)}=\left( \begin{array}{c} 0\\ 0\\1 \end{array} \right)
\label{locations-1}
\end{eqnarray}
This $\rho^{(1)}(\psi^{(1)})\/$ is invariant under an $U(2)(SU(2))\/$
subgroup of $SU(3)\/$. Exploiting this we can next transport $B^{\prime}\/$ to a
position $B^{(0)}\/$ while leaving $A^{(0)}\/$ fixed, and simultaneously
$C^{\prime}\/$ to some $C^{\prime \prime}\/$,  such that:
\begin{eqnarray}
A^{(0)},B^{\prime}, C^{\prime}
\shortstack[l]{{$U(2)$}\\{$\longrightarrow$}} A^{(0)},B^{(0)},C^{\prime
\prime }\quad : 
\nonumber \\
B^{(0)} \rightarrow \rho^{(2)}=\psi^{(2)}\psi^{(2) \dagger}
, \quad \psi^{(2)}=
\left( \begin{array}{c} 0\\ \sin{\xi}\\\cos{\xi} \end{array}
\right),\, 0<\xi< \frac{\pi}{2},
\label{locations-2}
\end{eqnarray}
thereby introducing the angle $\xi\/$. 
We have also secured that $(\psi^{(1)},\psi^{(2)})\/$ is real
positive. Now these two density matrices $\rho^{(1)}, \rho^{(2)}\/$
are invariant under a particular diagonal $U(1)\/$ subgroup of
$SU(3)\/$, whose elements are:
\begin{equation} 
d(\beta)={\rm diag} \left( e^{-2 i \beta},\
e^{ i \beta},e^{ i \beta}\right) \in SU(3),\, 0 \leq \beta < 2 \pi.
\label{u1-freedom}
\end{equation} 
We now have the freedom of transformations $d(\beta)\/$, which leave
$A^{(0)}\/$ and $B^{(0)}\/$ unchanged, to move $C^{\prime \prime}\/$ to a
convenient position $C^{(0)}\/$. A little thought shows that this can be
achieved as follows:
\begin{eqnarray}
&&A^{(0)}, B^{(0)}, C^{\prime \prime} 
\shortstack[l]{{$U(1)$}\\{$\longrightarrow$}}A^{(0)}, B^{(0)}, C^{(0)} \,:
\nonumber \\
&&C^{(0)} \rightarrow \psi^{(3)}=\left(\begin{array}{c} 
 \sin{\eta}\cos{\zeta} \\
 e^{i\chi_2}\sin{\eta}\sin{\zeta}  \\ 
\cos{\eta} \end{array} \right), 
\quad 0 < \eta < \frac{\pi}{2},\, 0 \leq \zeta \leq 
\frac{\pi}{2},\quad 0\leq
\chi_2 < 2 \pi.
\label{triangle}
\end{eqnarray}
We have parametrised $\psi^{(3)}\/$ in the manner of
eq.~(\ref{local-co-O}): the $U(1)\/$ freedom allows us to transform the
phase $\chi_1\/$ to zero, and an overall phase freedom 
has been used to make 
$(\psi^{(1)},\psi^{(3)})\/$ real positive. 
We now see that the description of the most general geodesic triangle in 
${\cal O}\/$, upto an overall $SU(3)\/$ transformation, involves the four
angle parameters $\xi, \eta, \zeta, \chi_2\/$. These are therefore
intrinsic to the shape and size of the triangle.
This counting agrees with the fact that $SU(3)\/$ is eight dimensional,
and choosing three points on ${\cal O}\/$ independently involves choosing
twelve independent coordinates.

Now we consider the geometric phase for the geodesic triangle  
$A^{(0)}, B^{(0)} ,C^{(0)} \/$. 
Using the result~(\ref{bargmann}) based on the Bargmann invariant, and the
conveniently chosen representative vectors in 
eqs.~(\ref{locations-1},~\ref{locations-2},~\ref{triangle}), we
find:
\begin{eqnarray}
\varphi_g[A^{(0)} B^{(0)} C^{(0)}]&=&
-{\rm arg}(\psi^{(1)}, \psi^{(2)})(\psi^{(2)}, \psi^{(3)})
(\psi^{(3)}, \psi^{(1)})
\nonumber \\
&=&+{\rm arg}(\psi^{(3)}, \psi^{(2)})
\nonumber \\
&=&{\rm arg}(\cos{\xi}\cos{\eta}+
\sin{\xi}\sin{\eta}\sin{\zeta} e^{-i \chi_2})
\nonumber \\
&=&{\rm arg}(1+
\tan{\xi}\tan{\eta}\sin{\zeta} e^{-i \chi_2})
\label{phig-3-level}
\end{eqnarray}
This is the generalisation of the Pancharatnam formula~(\ref{panch-solid}).
We see that the phase $\chi_2\/$ plays an important role: 
$\varphi_g[A^{(0)}B^{(0)} C^{(0)}]\/$ can be nonzero only if $\chi_2\/$
is nonzero. 

We can also express this geometric phase directly in terms of the vectors
${\bf n}^{(1)},\/{\bf n}^{(2)},\/{\bf n}^{(3)} \in 
{\cal O}\/$ corresponding to the
vertices $A^{(0)}, B^{(0)} ,C^{(0)} \/$ of the geodesic triangle:
\begin{eqnarray}
\varphi_g[A^{(0)}B^{(0)} C^{(0)}]&=&-\mbox{arg
Tr}(\rho^{(1)}\rho^{(2)}\rho^{(3)})
\nonumber \\
&=&-\tan^{-1} \left[\frac{2\sqrt{3}{\bf n}^{(1)}.{\bf n}^{(2)}\mbox{}_\wedge
{\bf n}^{(3)}}{({\bf n}^{(1)}+{\bf n}^{(2)}+{\bf n}^{(3)})^2 +2\/{\bf
n}^{(1)}.{\bf n}^{(2)} \star 
{\bf n}^{(3)} -\/2} \right].
\label{geo-phase-n123}
\end{eqnarray}
In this form the $SU(3)\/$ invariance is explicit.

We may collect our results of this and the previous Section to say:
given any geodesic triangle in the ray space ${\cal O}\/$ for
three-level systems, it is possible to find a piecewise constant
Hamiltonian to produce evolution along this triangle, such that the
dynamical phase vanishes, and the geometric phase is then given by the
intrinsic expressions~(\ref{phig-3-level},~\ref{geo-phase-n123}).  

\section{Concluding Remarks}
We have exploited the newly constructed extension of the 
Poincar$\stackrel{\prime}{\rm e}$ sphere
representation from two to three-level quantum systems, to develop a
generalisation of the Pancharatnam geometric phase formula for cyclic
evolution of such a system along a geodesic triangle in state space. We
have found that such a triangle is intrinsically defined by four angle
type parameters in contrast to the three needed for defining a triangle
on the Poincar$\stackrel{\prime}{\rm e}$ sphere $S^2\/$; and have obtained the explicit and simple
expression for the geometric phase in terms of them

It is easy to check that the original Pancharatnam
formula~(\ref{panch-solid})  emerges
form eq~(\ref{phig-3-level}) as a particular case. Namely if we take
$\zeta=\frac{\textstyle \pi}{\textstyle 2}\/$, the three Hilbert space vectors $
\psi^{(1)}, \psi^{(2)}, \psi^{(3)}\/$ of 
eqs~(\ref{locations-1},~\ref{locations-2},~\ref{triangle}) all lie in the two
dimensional $2-3\/$ subspace of ${\cal H}^{(3)}\/$, and involve just
three angle parameters $\xi , \eta, \chi_2\/$: 
\begin{equation}
\psi^{(1)}=\left(\begin{array}{c}0\\0\\1 \end{array}\right),\,\,
\psi^{(2)}=\left(\begin{array}{c}0\\\sin{\xi}\\\cos{\xi} \end{array}\right),\,\,
\psi^{(3)}=\left(\begin{array}{c}0\\e^{i\chi_2} \sin{\eta}\\\cos{\eta}
 \end{array}\right).
\end{equation}
Let the corresponding rays be represented by unit vectors ${\bf n}_1,
{\bf n}_2, {\bf n}_3\/$ on an $S^2\/$ and let the sides of the
corresponding spherical triangle be $a, b\/$ and $c\/$. Then from
eq~(\ref{rho-s2-1}), we have:

\begin{eqnarray}
{\bf n}_1. {\bf n}_2 &=& \cos{a} = 2\vert (\psi^{(1)},\psi^{(2)})\vert^2\
-1= \cos{2\xi}\,,
\nonumber \\
{\bf n}_1. {\bf n}_3 &=& \cos{b} = 2\vert (\psi^{(1)},\psi^{(3)})\vert^2\
-1= \cos{2\eta}\,,
\nonumber \\
{\bf n}_2. {\bf n}_3 &=& \cos{c} = 2\vert (\psi^{(2)},\psi^{(3)})\vert^2\
-1= 2 \vert\cos{\xi}\cos{\eta}+\sin{\xi}\sin{\eta}e^{i\chi_2}\vert^2 -1.
\label{inner-n_s}
\end{eqnarray}
Thus  $a=2\xi\/$, $b=2\eta\/$ and in the case of the angle $c\/$ we have 
\begin{equation}
\cos{\frac{c}{2}}=\vert\cos{\xi}\cos{\eta}+
\sin{\xi}\sin{\eta}e^{i\chi_2}\vert.
\end{equation}
Now from eq~(\ref{phig-3-level}) the geometric phase in this case is given by:
\begin{equation}
\varphi_g[A^{(0)} B^{(0)} C^{(0)}]=arg(
\cos{\xi}\cos{\eta}+\sin{\xi}\sin{\eta}e^{-i\chi_2}),
\end{equation}
or equally well (apart from the sign) by:
\begin{equation}
\cos(\varphi_g[A^{(0)} B^{(0)} C^{(0)}])=\frac{
\cos{\xi}\cos{\eta}+\sin{\xi}\sin{\eta}\cos{\chi_2}}{
\vert\cos{\xi}\cos{\eta}+\sin{\xi}\sin{\eta}e^{i\chi_2}\vert}.
\end{equation}
Focussing on the $\chi_2\/$ dependence here (and as we have seen earlier
this is the crucial aspect), and using eq.~(\ref{inner-n_s}) we have:
\begin{equation}
\cos{\xi}\cos{\eta}+\sin{\xi}\sin{\eta}\cos{\chi_2}=
\frac{(1+\cos{a}+\cos{b}+\cos{c} )}{4 \cos{a/2} \cos{b/2}},
\end{equation} 
leading to
\begin{equation}
\cos(\varphi_g[A^{(0)} B^{(0)} C^{(0)}])=
\frac{(1+\cos{a}+\cos{b}+\cos{c} )}{4 \cos{a/2} \cos{b/2} \cos{c/2}}.
\end{equation}
However the right hand side here is precisely $\cos{\frac{1}{2}
\triangle (a,b,c)}\/$, where $\triangle (a,b,c)\/$ is the solid angle
subtended at the origin of $S^2\/$ by the spherical triangle with sides 
$a,b$ and $c\/$. In this way the Pancharatnam result  $
\varphi_g[A^{(0)} B^{(0)} C^{(0)}]=\frac{1}{2}
\triangle(a,b,c)\/$ is recovered. (The sign can also be recovered with
some additional effort).

This same verification  leads us to the  following significant remark:
as long as one is interested in cyclic evolution of {\em any} quantum system
along a geodesic {\em triangle} in ray space, however large the
dimension of the Hilbert space ${\cal H}\/$ may be,
our result~(\ref{phig-3-level})
for the geometric phase is applicable and is completely general. This is
because a triangle involves (at most) three independent vectors
$\psi^{(1)}, \psi^{(2)}, \psi^{(3)} \in {\cal H}\/$, and these always lie
in some three dimensional subspace of ${\cal H}\/$. Thus nothing
additional is needed to handle geometric phases for cyclic evolutions
along geodesic triangles for $N$-level systems, for any $N \geq 4\/$.
In this sense the original Pancharatnam result~(\ref{panch-solid}) deals with a
degenerate case, since it is concerned with evolution along a triangle
but for a two-level system.  There is then no place for the additional
angle variable $\zeta\/$ to appear as it does in the fully general
formula~(\ref{phig-3-level}).

A discussion of feasible experimental schemes to check the validity of
our main result~(\ref{phig-3-level}), 
at least in some nontrivial cases which do go
beyond the two-level situation, will be the subject of a forthcoming
publication.  
 

{\large \bf Acknowledgments}\\
Arvind thanks University Grants Commission India for financial
support and KSM thanks the JNCASR for Visiting Fellowship and the CTS for
providing facilities.

\setcounter{figure}{0}

\begin{references}
\bibitem {local-var} The choice of these variables differs from that in
refs~\cite{su-3}.
\end{references}
\end{document}